# Extragalactic Background Light and Gamma-Ray Attenuation

Joel R. Primack[a], Alberto Domínguez[a,b,c], Rudy C. Gilmore[a,d], and Rachel S. Somerville[e,f]

[a]*Physics Department, University of California, Santa Cruz, CA 95064 USA*
[b]*Instituto de Astrofisica de Andalucia, CSIC, Apdo. Correos 3004, E-18080 Granada, Spain*
[c]*Departamento de Fısica Atomica, Molecular y Nuclear, Universidad de Sevilla, Apdo. Correos 1065, E-41080 Sevilla, Spain*
[d]*Scuola Internazionale Superiore di Studi Avanzati (SISSA), Via Bonomea 265, 34136, Trieste, Italy*
[e]*Space Telescope Science Institute, 3700 San Martin Drive, Baltimore, MD 21218*
[f]*Department of Physics and Astronomy, Johns Hopkins University, Baltimore, MD 21218*

**Abstract.** Data from (non-) attenuation of gamma rays from active galactic nuclei (AGN) and gamma ray bursts (GRBs) give upper limits on the extragalactic background light (EBL) from the UV to the mid-IR that are only a little above the lower limits from observed galaxies. These upper limits now rule out some EBL models and purported observations, with improved data likely to provide even stronger constraints. We present EBL calculations both based on multiwavelength observations of thousands of galaxies and also based on semi-analytic models, and show that they are consistent with these lower limits from observed galaxies and with the gamma-ray upper limit constraints. Such comparisons "close the loop" on cosmological galaxy formation models, since they account for all the light, including that from galaxies too faint to see. We compare our results with those of other recent works, and discuss the implications of these new EBL calculations for gamma ray attenuation. Catching a few GRBs with ground-based atmospheric Cherenkov Telescope (ACT) arrays or water Cherenkov detectors could provide important new constraints on the high-redshift star formation history of the universe.

## INTRODUCTION

The EBL is very difficult to observe directly because of foregrounds from the Milky Way and the solar system, especially the zodiacal light [1]. Reliable lower limits are obtained by integrating the light from observed galaxies. The best upper limits come from (non-) attenuation of gamma rays from distant blazars due to scattering on the EBL producing $e^+e^-$ pairs, but these upper limits are uncertain because of the unknown emitted spectrum of these blazars. This paper concerns both the optical-IR EBL relevant to attenuation of TeV gamma rays, and also the UV EBL relevant to attenuation of multi-GeV gamma rays from very distant GRBs & blazars observed by Fermi and low-threshold ground-based ACTs, including the larger future array CTA and water Cherenkov detector HAWC. Just as IR light penetrates dust better than shorter wavelengths, so lower energy gamma rays penetrate the EBL better than higher energy. Detectors with low energy thresholds are therefore essential to see high-redshift gamma rays.

If we know the intrinsic spectrum of a source of gamma rays at redshift $z$, we can use $dN/dE|_{obs} = \exp[-\tau(E,z)]\, dN/dE|_{int}$ to infer the optical depth $\tau(E,z)$ due to $\gamma\gamma \to e^+e^-$ from the observed spectrum. In practice, we typically *assume* that $dN/dE|_{int}$ is not harder than $E^{-\Gamma}$ with $\Gamma = 1.5$, since local sources have $\Gamma > 2$. Figure 1 summarizes upper limits using this method (curves with downward arrows—the curve labeled "extreme" assumes $\Gamma = 2/3$), direct measurements of the EBL (open symbols), lower limits from observations of galaxies (filled symbols).

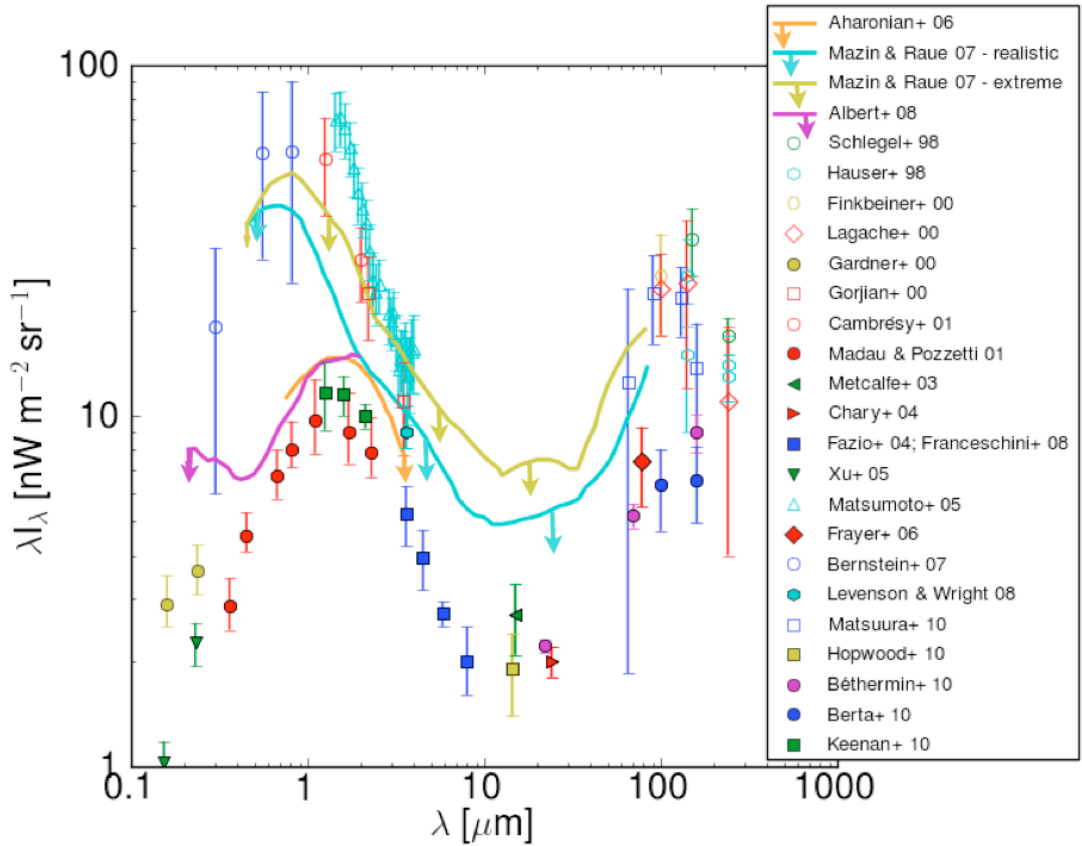

**FIGURE 1.** Observations of the extragalactic background light (EBL): direct observations (open symbols), lower limits from deep galaxy observations (filled symbols), and upper limits (curves with arrows). For references summarized at right see [2].

There are basically four approaches to calculate the EBL: (1) *Backward Evolution*, which starts with the existing galaxy population and scales it backward in time as a power-law in $(1 + z)$ (e.g., [3]). This is not well justified, since the high-redshift galaxy population emits radiation with a rather different spectral energy distribution (SED) than the nearby population. (2) *Backward Evolution Inferred from Observations*, attempting to correct for the changing luminosity functions and SEDs with redshift and galaxy type (e.g., [4-6]). (3) *Evolution Directly Observed and Extrapolated* based on a large set of multiwavelength galaxy observations [2], discussed below and in more detail by Alberto Domínguez in a Parallel Session at the

Texas 2010 conference [7]. (4) *Forward Evolution*, which begins with cosmological initial conditions, and semi-analytically models gas cooling in dark matter halos, formation of galaxies including stars and AGN, feedback from these phenomena, stellar evolution, and scattering, absorption, and re-emission of light by dust ([8,9], discussed below).

In this paper we discuss the EBL calculated using methods (3) and (4) and compare with some recent work based on method (2). We then discuss the implications for gamma-ray attenuation, including the Fermi Gamma Ray Space Telescope and existing and planned arrays of ground-based gamma-ray telescopes. Throughout this paper we use a standard $\Lambda$CDM cosmology with matter density $\Omega_m = 0.3$, cosmological constant density $\Omega_\Lambda = 0.7$, and Hubble parameter $H_0 = 70$ km/s/Mpc.

## EBL CALCULATED DIRECTLY FROM OBSERVATIONS

We have recently developed a new method to calculate the EBL [2], using a large set of multiwavelength galaxy data from the All-wavelength Extended Groth strip International Survey (AEGIS)[1]. We use local ($z < 0.2$) data from the Sloan Digital Sky Survey (SDSS) to determine the local contribution to the EBL of different types of galaxies. For redshifts $0.2 < z < 1$ we fit the multiwavelength SEDs of ~6000 AEGIS galaxies to the SWIRE templates[2], propagating errors in the SED measurements and in the fits; and we used this to determine the mix of SED-types as a function of redshift. We showed that this data set is complete enough to allow us to do this. At higher redshifts we use different extrapolations to bound our ignorance. We use the evolving K-band luminosity function [10] to normalize our EBL calculation. Our resulting fiducial EBL is shown as the red long-short dashed curve in Fig. 2, with the pink band showing all the uncertainties including those from the extrapolations to higher redshifts. This is the first determination of the evolving EBL from 0.1 to 1000 μm directly from observed luminosity functions and SEDs. In this paper we also discuss the implications for gamma-ray attenuation, and we have made detailed EBL and attenuation data available for download.[3] Figure 2 also displays the EBL results from our new semi-analytic model (SAM) of the evolving galaxy population, discussed below, and shows that they are generally consistent with our observational determination of the EBL except at the longest wavelengths.

## FORWARD EVOLUTION CALCULATION OF THE EBL

Semi-analytic models (reviewed in [11]) have been shown to reproduce many observed properties of galaxies, and they remain the best method for modeling the properties of the evolving population of galaxies in the universe. We developed some of the first SAMs that included dust absorption of light [12], and we have for some time used this approach to model the EBL [13-15]. Our new SAM [8] is based on the standard $\Lambda$CDM hierarchical formation of dark matter halos, and it models galaxy formation using a set of recipes for gas accretion and cooling, star formation, stellar feedback, chemical enrichment, black hole growth and AGN feedback. The evolving stellar population of these galaxies is used to predict their evolving unattenuated SEDs. We use simple analytic recipes describing the absorption and re-emission of starlight by dust in the interstellar medium of galaxies to predict galaxy counts and luminosity functions from the far-ultraviolet to the sub-mm from redshift 5 to the present, and compare with an extensive compilation of observations [8,9].

---
[1] http://aegis/ucolick.org/
[2] http://www.iasf-milano.inaf.it/~polletta/templates/swire_templates.html
[3] http://side.iaa.es/side/EBL/

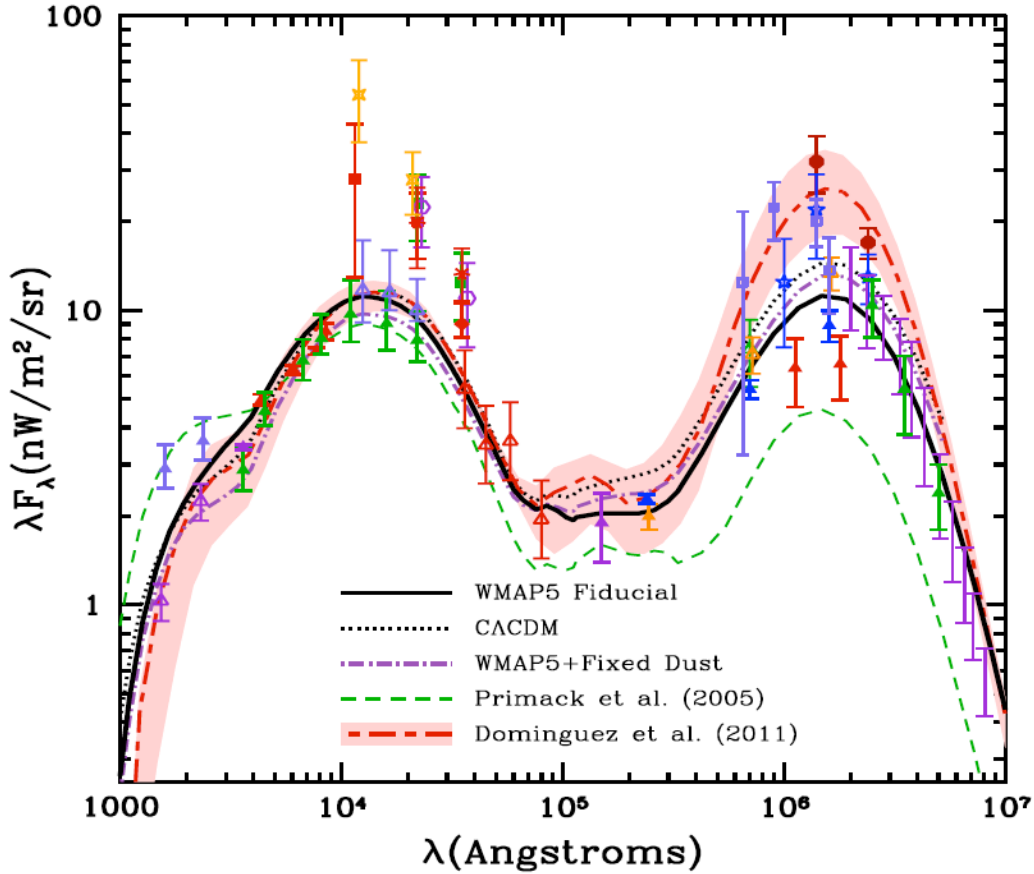

**FIGURE 2.** The EBL in the local universe compared with observations. The solid black curve is our fiducial SAM, based on WMAP5 cosmological parameters with an evolving dust attenuation model, and the purple dot-dashed curve is the same SAM with a non-evolving dust model. The red short-long-dashed curve is the result of our observational determination of the EBL [2], and the pink band shows the effect of uncertainties in the SED photometry and template fits, the K-band luminosity function used for normalization, and the extrapolation of SED-types to redshifts $z > 1$. Two of our older models are shown for reference, our 2005 EBL model [14] (dotted), and the CΛCDM EBL from a 2008 version of our SAM [16] that used WMAP1 cosmological parameters with a non-evolving dust model (green dashed curve). (This is Fig. 4 of [9], which gives complete references to the observations plotted.)

We find that in order to reproduce the observed rest-UV and optical luminosity functions at high redshift, we must assume an evolving normalization in the dust-to-metal ratio, implying that galaxies of a given bolometric luminosity (or metal column density) must be less extinguished than their local counterparts. In our model, all energy absorbed by dust is re-emitted at longer wavelengths, using dust emission templates based on Spitzer observations [16]; the total luminosity of each model galaxy is used to select the appropriate IR emission template. This has been shown [17] to work as well in this context as using the full radiative transfer model GRASIL [18]. In our fiducial model, we find remarkably good agreement with observations from rest 1500 Å to 250 μm, as shown for example in Fig. 3. At longer wavelengths, most dramatically in the sub-mm, our models underpredict the number of bright galaxies by a large factor. Despite this, the good agreement with observations implies that we can use our fiducial SAM to calculate the EBL reliably except at the very far IR.

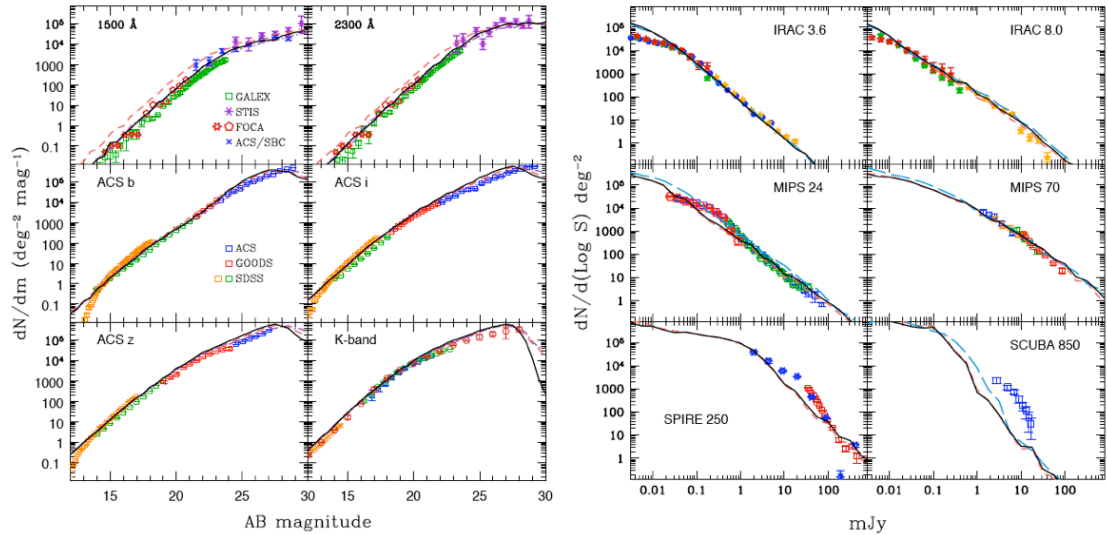

**FIGURE 3.** (a) Number counts in the GALEX UV bands and the four HST ACS bands, compared with our SAM predictions. (b) Number counts from four Spitzer (IRAC and MIPS) bands, Herschel 250 μm and SCUBA 850 μm, compared with our SAM predictions. Our SAM predictions are WMAP5 Fiducial (solid black), WMAP5 single component (Calzetti) dust model (red dashed), WMAP5 with fixed dust absorption and emission templates from [20] (long-dashed blue). The only clear failures are for Herschel 250 μm and SCUBA 850 μm. (Figures 14 and 15 from [8].)

From our WMAP5 Fiducial model of the evolving galaxy population, we calculate the evolution of the luminosity density of the universe, shown in Fig. 4.

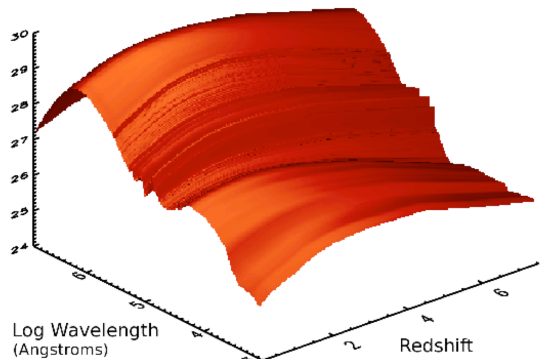

**FIGURE 4.** Three-dimensional representation of the evolving luminosity density (the vertical axis is erg/Hz/s/Mpc$^3$) in our WMAP5 Fiducial model as a function of $\lambda$ and $z$. (Figure 2 of [9].)

The evolution of the EBL with redshift is shown graphically in Fig. 5, in two ways: in physical and co-moving coordinates. The left panel shows that the EBL was much higher in the past, especially in the optical and near-IR and in the far-IR. The right panel shows how the present-day EBL was generated as a function of redshift. This EBL evolution must be taken into account in calculating attenuation of gamma rays from all but the nearest extragalactic sources. The change in the functional form of the EBL means that a simple $z$-dependent scaling model is inadequate.

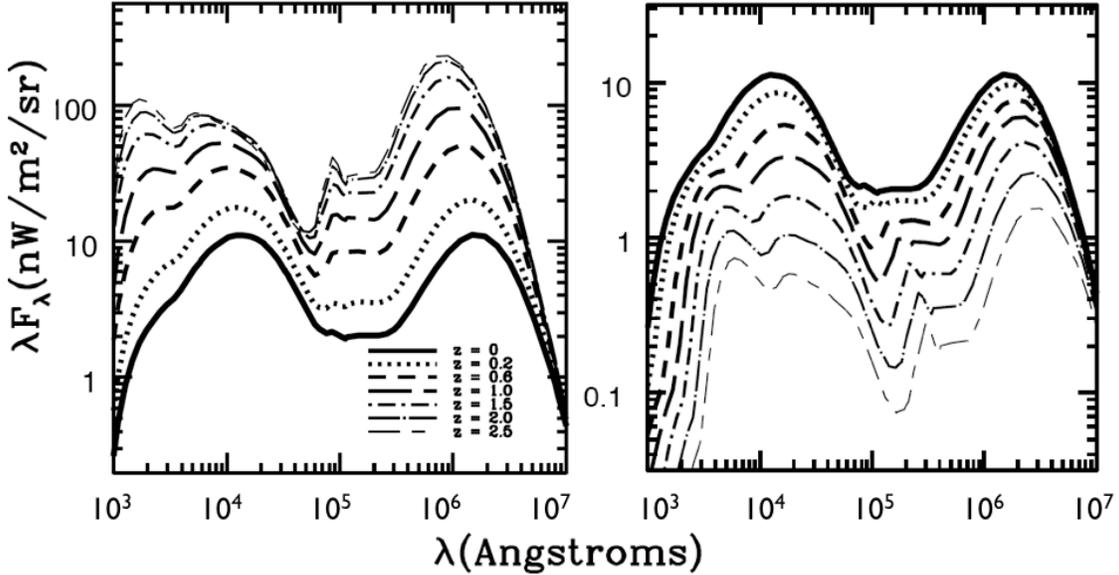

**FIGURE 5.** The evolution of the EBL in our WMAP5 Fiducial model. This is plotted on the left panel in standard units. The right panel shows the build-up of the present-day EBL by plotting the same quantities in comoving units. The redshifts from 0 to 2.5 are shown by the different line types in the key in the left panel. (From Fig. 5 of [9].)

## GAMMA RAY ATTENUATION

Gamma ray attenuation due to $\gamma\gamma \rightarrow e^+e^-$ is calculated by integrating the cross section times the proper density of background photons along the line of sight to the emitting redshift, and integrating over the scattering angle $\theta$, where $\theta = \pi$ corresponds to a head-on collision. The most probable scattering angle is $\theta \approx \pi/2$. If we assume $\theta = \pi/2$, then the characteristic wavelength $\lambda_{bg}$ of the background photons that will most strongly affect a gamma ray of energy $E_\gamma$ is given by $\lambda_{bg} = 1.2\,(E_\gamma/\text{TeV})\,\mu\text{m}$.

We have calculated gamma-ray attenuation as a function of the redshift of the source and the observed gamma-ray energy, from the evolving EBL determined both observationally and from our SAM calculations. This is shown in the left panel of Fig. 6.

A more general way to show the EBL attenuation is to plot the "Attenuation Edge" redshift where the optical depth $\tau$ reaches a certain value as a function of gamma-ray energy, which is presented in the right panel of Fig. 6 out to redshift 5 for $\tau = 1, 3,$ and

10. This figure shows that gamma-ray telescopes with lower threshold energies will allow us to peer more deeply into the universe. In [19] we did an improved calculation for lower gamma-ray energies using a range of models of the hard-UV EBL evolution that exemplify the current range of uncertainty regarding the sources of ionizing radiation and taking into account the optical depth of the universe to ionizing radiation.

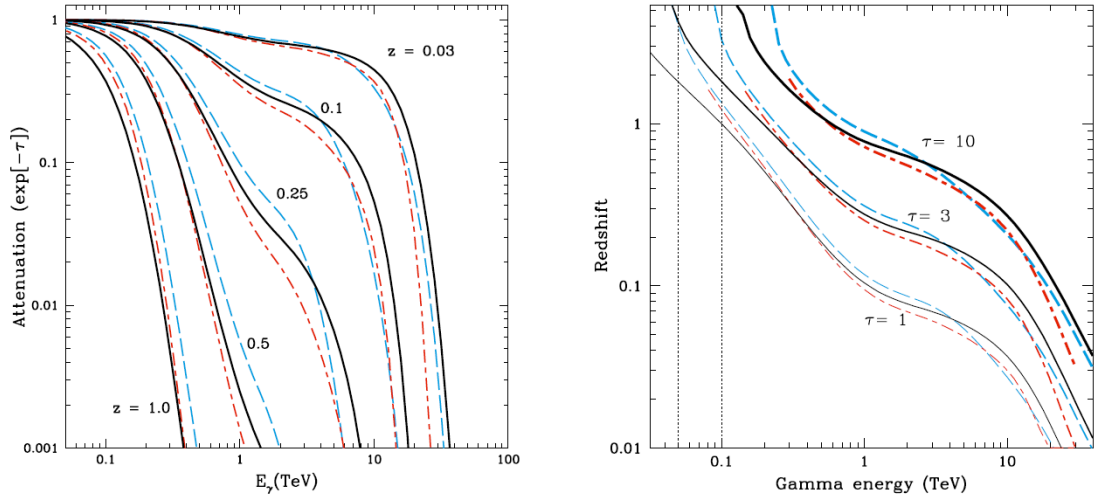

**FIGURE 6.** (*Left Panel*) Gamma ray attenuation for gamma rays of observed energy $E_\gamma$ for sources at redshifts 0.03, 0.1, 0.25, 0.5, and 1, for our observational EBL (red long-short dash), our WMAP5 Fiducial model (solid), and our WMAP5 SAM with a fixed dust model using the pre-Spitzer dust templates [20] that we used in earlier EBL calculations. The plateau between 1 and 10 TeV at low redshifts is a consequence of the mid-IR velley in the EBL spectrum. (*Right Panel*) Gamma Ray Attenuation Edge for the same models. The curves show the redshift at which the pair-production optical depth $\tau$ reaches the value 1, 3, or 10 as a function of observed gamma ray energy. We have included thin lines to guide the eye at 50 and 100 GeV. (Figures 8 and 9 of [9].)

Both our observationally determined EBL (taking into account the uncertainties) and the EBL from our WMAP5 Fiducial SAM calculation are consistent with the lower limits from galaxy observations and the upper limits from non-attenuation of gamma-rays from distant blazars shown in Fig. 1. In Fig. 7 we show that this is true for the two highest-redshift blazars detected by the MAGIC atmospheric Cherenkov telescope (ACT), and for known blazars from all ACTs.

A detailed analysis of the EBL constraints available from all Fermi observations of blazars and GRBs was the subject of a recent paper by the Fermi collaboration [21]. These limits do not constrain the UV flux in our EBL models [2,8,9,19]. However, the EBL models of [3] are ruled out at the 5σ level.

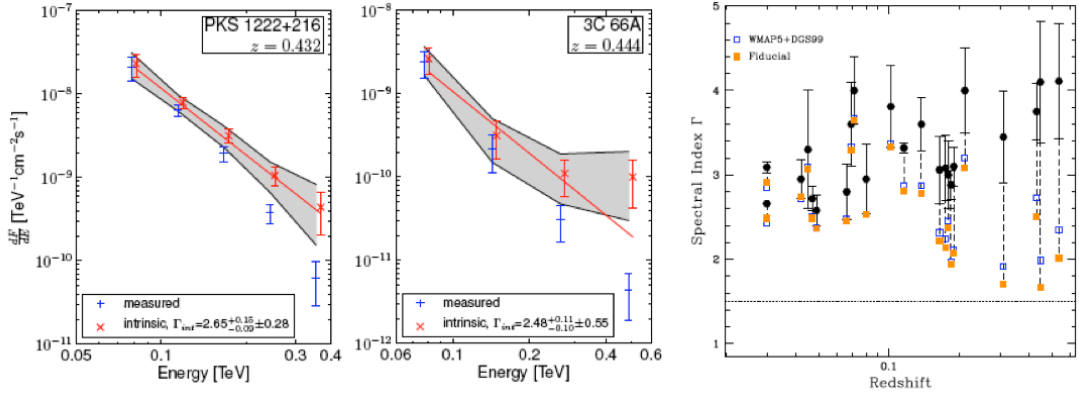

**FIGURE 7.** Implications of our EBL models for reconstruction of the unattenuated index $\Gamma_{int}$ of blazar gamma-ray spectra. Panels (a) and (b) show that $\Gamma_{int} > 1.5$, consistent with expectations, for the two highest-redshift MAGIC blazars, and panel (c) shows that this is true for all the high-redshift blazars for our WMAP5 models, both Fiducial and with fixed dust using the templates from [20]. (Panels (a) and (b) are from [7] and panel (c) is from [9].)

## COMPARISON WITH OTHER WORKS

Comparisons with other determinations of the EBL are given in [2], and other methodologies and results are discussed at considerable length in section 4 of [9]. Here we compare our results graphically with those of other authors in Fig. 8. Our EBL models are seen to evolve similarly to the Franceschini et al. model [5] out at least to redshift 1. The "best fit" EBL of Kneiske et al. [4] is considerably higher than any of our models, probably largely because of the very different evolution of the star formation rate density assumed, peaking at $z = 1.2$ while ours peaks at higher $z$. The update to this model in [6] is more similar to our WMAP5 Fiducial model. The approach is similar in the Finke et al. model [22], but their assumed stellar initial mass function (IMF) produces slightly more high-mass stars than the Chabrier IMF assumed in our SAM, and this model has a slightly higher normalization in the optical and near-IR. Their SED in the mid- and far-IR is different from ours; like Kneiske et al. they assume a simplified dust emission model based on thermal black body emission at three fixed temperatures while we use observational templates based on the galaxies' total bolometric luminosity. The Stecker et al. models [3], in which the local galaxy population just grows brighter back in time, give EBLs considerably higher than ours in the optical and near-IR, and with a different spectral shape. As mentioned above, these models are in conflict with the Fermi limits [21].

At nearly all wavelengths considered, our observational and SAM Fiducial EBL models are near the levels of flux resolved in discrete background counts, and we find agreement with the claim of Madau and Pozetti [23] and Keenan et al. [24] that nearly all the light in the optical and near-IR EBL peak is produced by resolved galaxies. Our EBL models lie below the claimed direct detections in Fig. 1, but we note the large statistical and systematic errors on these measurements, and the fact that they disagree with the upper limits from non-attenuation of gamma rays from blazars.

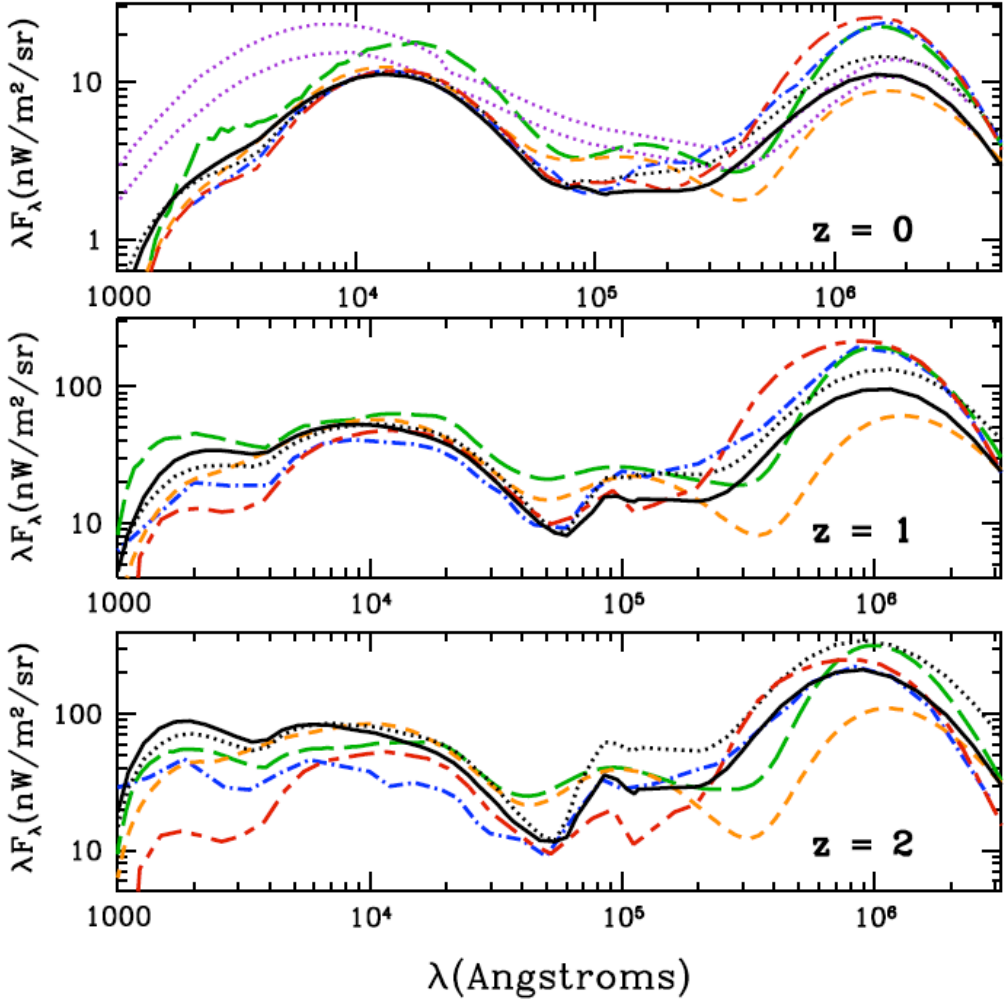

**FIGURE 8.** Our EBL predictions compared with several models from the literature. Long-short dashed red curves show the proper flux density from our observational EBL determination [2]; solid and dotted black lines show the proper flux density from our WMAP5 and CΛCDM models. Results are shown in the local universe and at $z = 1$ and $z = 2$. Other lines are from Franceschini et al. [5] (dashed-dotted blue), the best-fit model of Kneiske et al. (2004) [4] (long-dashed green), and model 'C' from Finke et al. (2010) [23] (dashed orange). The baseline and fast evolution models of [3] are the low and high dotted violet points in the $z = 0$ panel. (Figure 12 of [9].)

Using a novel method involving searching for breaks in the gamma-ray spectra of nearby blazars potentially caused by attenuation by the EBL, Orr et al. [25] have recently claimed to detect a 2× to 1.6× higher local EBL than our observational determination [2] from about 0.4 to 3 μm, and a somewhat lower EBL than we found at 15 μm. They claim that these new results are inconsistent at the 3σ level with our observational determination of the EBL. While we are intrigued by the possibility to get lower as well as upper limits from blazar spectra, we do not find this claim to be convincing. We found using statistical F-tests that only two of the blazar spectra considered in [25], 1ES1101-232 and RGBJ0152+017, are better fit by a broken power law, a condition that must be satisfied for the validity of their methodology. Increasing the rather optimistic systematic uncertainty of 0.1 in the measured blazar spectral indices assumed in [25] would lessen the discrepancy between their result and direct determinations of the EBL and the upper limits shown by the lower curves with downward pointing arrows in Fig. 1. If, however, it were really true that there is a component of the EBL that cannot be detected from the faintest observed galaxies including reasonable incompleteness corrections, that would presumably imply that the source is faint galaxies at very high redshifts. There are rather strong physical arguments against this possibility (e.g., [26]). We note that if such an extra component of the EBL from high redshifts were really present, it would imply very significant attenuation of gamma rays from high-redshift GRBs – which could be checked by detecting enough photons from such GRBs to determine their spectra.

## CATCHING GRB GAMMA-RAYS WITH FERMI AND ACTS

The powerful arrays of atmospheric Cherenkov telescopes H.E.S.S., MAGIC, and VERITAS have begun operating during the past decade, and the Swift satellite is capable of localizing GRBs within seconds. However, despite major campaigns to respond to satellite GRB alerts, no conclusive detection of a GRB with an ACT has yet been made. Since the launch of the Fermi satellite in June 2008, its Large Area Telescope (LAT) has detected GRBs at with $E_\gamma > 100$ MeV at a rate of about 10/yr. The LAT GRB with highest determined redshift thus far is GRB 080916C at $z = 4.24$, with a rest-frame energy of $E_{\gamma\text{-rest}} = 69$ GeV. GRB 090902B at $z = 1.82$ had the highest rest-frame energy yet detected, $E_{\gamma\text{-rest}} = 94$ GeV.

In a recent paper [27], we modeled the probability of detecting GeV gamma rays from GRBs with the Fermi LAT and the MAGIC ACT. Using simple assumptions about the GRB rate and a GRB spectral extrapolation based on the BATSE and Swift GRB data, this paper predicted the number of GRBs that could be detected per year and the likely number of photons above background for each GRB. The Fermi LAT was predicted to have a detection rate of 3 or 4 GRBs above 10 GeV per year, in line with the first-year observations, but higher than the rate since then. The detection rate for MAGIC was predicted to be about 0.2-0.3 events per year, taking into account the helpful fact that Swift finds more GRBs on the night side of the earth and using the somewhat optimistic design assumptions about the MAGIC telescope response time to burst triggers. Despite the low predicted rate for MAGIC, the potential result of a GRB detection near zenith was found to be 10s to 1000s of counts in the lower energy range of the experiment, with a potentially large scientific payoff.

We are now finishing an improved calculation [28] with a more sophisticated treatment of the spectral extrapolation to high energies and of the telescope detection capabilities, including the proposed new large arrays of large ACTs such as CTA [29], which with larger effective areas and lower energy thresholds could detect even more photons from a GRBs at a wider redshift range. Rudy Gilmore presented our preliminary results at a parallel session of the Texas 2010 meeting [30].

In [27] we extrapolated GRB emission to high energies assuming that the flux at VHE energies could be described as a fixed fraction of the flux at lower energies, namely Flux(0.1-5 GeV) = 0.1 Flux(0.02-2 MeV), reflecting the typical ratio inferred from coincident BATSE-EGRET observations. In [28] we now use LAT GRB observations to improve our flux estimates. We consider two different models, a fixed flux ratio model like that used in [27], and an alternative model in which the higher energy flux is a smooth extrapolation of the Band spectra determined at lower energy. We assume that the GRB redshift distribution is that of the Swift GRBs (which have been detected at a rate of about 95/yr), and we calculate gamma-ray attenuation using the fiducial model of [19]. Because of the strong attenuation of higher-energy gamma rays, we find that a low ACT array threshold energy, such as the 20 GeV energy threshold for CTA discussed in [30], significantly increases the redshift range and the number of GRB gamma rays detected. In calculating the effect of the time delay between the Swift GRB detection and the ACT availability we have assumed that the afterglow declines as $t^{-1.5}$, as has been found for a number of LAT GRBs [31]. Part of the point of these calculations is to help optimize the CTA design. We plan to do similar calculations when we learn the proposed design parameters of the High Altitude Water Cherenkov (HAWC) detector.

## ACKNOWLEDGMENTS

JRP, AD, and RCG acknowledge support from a Fermi Guest Investigator Grant. Relevant research of JRP and RCG was also supported by NASA ATP grant NNX07AGG4G and NSF-AST-1010033. RCG was supported during this work by a SISSA postdoctoral fellowship. AD's research has been supported by the Spanish Ministerio de Educación y Ciencia, the Eurpean regional development fund (FEDER) under projects FIS2008-04189 and CPAN-Ingenio (CSD2007-00042), and by the Junta de Andalalucía (P07-FQM-02894). AD, RCG, and RSS thank UCSC for hospitality during many visits.